\documentclass[fleqn,usenatbib]{mnras}

\usepackage{newtxtext,newtxmath}

\usepackage[T1]{fontenc}
\usepackage{ae,aecompl}

\usepackage{graphicx}	
\usepackage{amsmath}	

\newcommand{\bea}{\begin{eqnarray}}
\newcommand{\eea}{\end{eqnarray}}

\title[Annual cycle of J1726+0639]{The Annual Cycle in Scintillation Timescale of PMN~J1726+0639}

\author[H. E. Bignall et al.]{Hayley E. Bignall,$^{1,4}$\thanks{E-mail: Hayley.Bignall@manlyastrophysics.org}
Artem V. Tuntsov,$^{1}$\thanks{E-mail: Artem.Tuntsov@manlyastrophysics.org}
Jamie Stevens,$^{2}$
Keith Bannister,$^{3}$
\newauthor
Mark A. Walker,$^{1}$
Cormac Reynolds.$^{4}$
\\
$^{1}$Manly Astrophysics, 15/41-42 East Esplanade, Manly, NSW 2095, Australia\\
$^{2}$CSIRO Paul Wild Observatory, 1828 Yarrie Lake Road, Narrabri, NSW 2390, Australia\\
$^{3}$CSIRO Astronomy and Space Science, PO Box 76, Epping NSW 1710, Australia\\
$^{4}$CSIRO Astronomy and Space Science, PO Box 1130, Bentley WA 6102, Australia\\
}

\date{}

\pubyear{2021}

\begin{document}
\label{firstpage}
\pagerange{\pageref{firstpage}--\pageref{lastpage}}
\maketitle

\begin{abstract}
We discovered rapid intra-day variability in radio source PMN~J1726+0639 at GHz frequencies, during a survey to search for such variability with the Australia Telescope Compact Array. Follow-up observations were conducted over two years and revealed a clear, repeating annual cycle in the rate, or characteristic timescale, of variability, showing that the observed variations can be attributed to scintillations from interstellar plasma inhomogeneities. The strong annual cycle includes an apparent "standstill" in April and another in September. We fit kinematic models to the data, allowing for finite anisotropy in the scintillation pattern. 
The cycle implies a very high degree of anisotropy, with an axial ratio of at least $13:1$, and the fit is consistent with a purely one-dimensional scintillation pattern. 
The position angle of the anisotropy, and the transverse velocity component are tightly constrained. The parameters are inconsistent with expectations from a previously proposed model of scattering associated with plasma filaments radially oriented around hot stars. We note that evidence for a foreground interstellar cloud causing anomalous Ca~{\sc ii} absorption towards the nearby star Rasalhague ($\alpha$~Oph) has been previously reported, and we speculate that the interstellar scintillation of PMN~J1726+0639 might be associated with this nearby cloud.
\end{abstract}

\begin{keywords}
ISM: general -- ISM: structure -- radio continuum: galaxies -- radio continuum: transients -- circumstellar matter
\end{keywords}


\section{Introduction}
\label{section:intro}

Sufficiently compact (sub-milliarcsecond-scale), centimetre-wave radio sources scintillate, as a result of inhomogeneities in the ionized interstellar medium (ISM) \citep[e.g.][]{Lovell08}. Scintillation can be thought of as a spatial flux density pattern -- i.e. the source projected through a transparent plasma ``screen'' -- that drifts relative to the Earth. For sources at cosmological distances, the velocity of the pattern is essentially that of the screen \citep{effectivevelocity}, and therefore the change in the velocity of the Earth as it orbits the Sun can strongly affect the characteristic variability timescales. Such an annual modulation has been reported for more than a dozen sources (\citealt{cycle1819}, \citealt{jaunceymacquart}, \citealt{rickettetal}, \citealt{Bignall19}, and references therein; \citealt{Said20}; \citealt{Oosterloo20}; \citealt{Wang21}). 
For many of these, the annual cycles have been shown to be consistent with highly anisotropic, essentially one-dimensional scattering \citep[e.g.][]{walkerdebruynbignall,Bignall19,Oosterloo20,Wang21}, and the orientation of the respective anisotropy axes along with the projected velocity of the screen were determined. 

Interstellar scintillations of extragalactic sources (generally active galactic nuclei, AGN) on timescales of less than an hour are rarely observed, whereas \citet{Lovell08} showed that slower, intra- and inter-day, variations are displayed by a large fraction of compact sources. Broad-band intra-hour variability (IHV; also used to denote an "intra-hour variable" source) due to weak or refractive scintillation implies scattering within tens of parsec of the observer \citep[e.g.][]{cycle1819}. Such nearby scattering offers the possibility to reliably determine the nature and origin of the scattering material. The more distant structures responsible for slower scintillations are generally more difficult to precisely locate, due to the complexity of the ionised ISM along the line-of-sight, and the assumption of variations dominated by a single, "thin" scattering screen may not hold in these cases.  

An earlier suggestion that the scattering plasma in the foreground of J1819+3845 might be associated with the nearby A star Vega \citep{twostation1819}, along with the recent discovery of another IHV source, PKS~B1322$-$110 \citep{Bignall19}, just 8.5 arcminutes away from Spica, the Sun's closest B star neighbour, prompted \citet{walkeretal2017} to examine possible connections between the known intra-hour variables and hot stars. 
A conclusion of that study was that the scintillations of both J1819+3845 and PKS~1257$-$326 are due to plasma associated with nearby A stars --- the stars being Vega in the case of J1819+3845, and Alhakim ($\iota$ Cen), in the case of PKS~B1257$-$326. The picture that was suggested by \citet{walkeretal2017} is of plasma filaments that are radially oriented around the host star, and co-moving with it. The annual cycle subsequently observed for PKS~B1322$-$110 supports this picture, albeit with slightly more ambiguity in the fitted parameters, compared with the two aforementioned sources, due to the low ecliptic latitude of PKS~B1322$-$110. The estimated probability of chance association of Spica with the scintillation of PKS~B1322$-$110 is $\sim 1\%$ \citep{Bignall19}, whereas for the two earlier established cases, the probability of chance associations is $\sim10^{-4}$ \citep{walkeretal2017}. 

Until recently, J1819+3845, PKS~B1257$-$326 and PKS~B1322$-$110 were the only IHV sources with well-studied annual cycles. The advent of widefield cm-wavelength telescopes (namely, Apertif at Westerbork, and ASKAP) has led to the discovery of multiple intra-hour variable sources at frequencies near $\sim 1$\,GHz \citep{Oosterloo20,Wang21}. IHV at such low frequencies implies scattering within a few parsec of the observer. The implications of these recent discoveries are discussed further in Section~\ref{section:discussion}.

To test the proposed association between interstellar scintillation and plasma around hot foreground stars, we undertook a survey with the Australia Telscope Compact Array to search for more intra-hour scintillating AGN among 506 compact radio sources. %
Among them, 
PMN~J1726+0639 \citep{Griffith95} was found to show large intra-hour variations in its flux density. 
In this paper we report the results of tracking the rate of flux density variations in PMN~J1726+0639 for just over two years, from February 2018 to March 2020. In Section~\ref{section:data} we describe the observations and extraction of flux density measurements.  Our method of inferring variability rates in Section~\ref{section:analysis} allows us to characterise the scintillation rate during slow phases of the annual cycle, where traditional methods of analysis struggle. We fit the kinematic parameters of the annual cycle to the data, for the case assuming totally anisotropic scattering, as well as the more general case of finitely anisotropic scattering (i.e.\ an elliptical characteristic scintillation pattern). In Section~\ref{section:discussion} we compare the results to the predictions of the model connecting the scintillations with nearby hot stars, and discuss the implications. Conclusions are presented in Section~\ref{section:conclusions}.

\section{Observations and Data reduction}
\label{section:data}

\begin{figure}
	\includegraphics[width=\columnwidth]{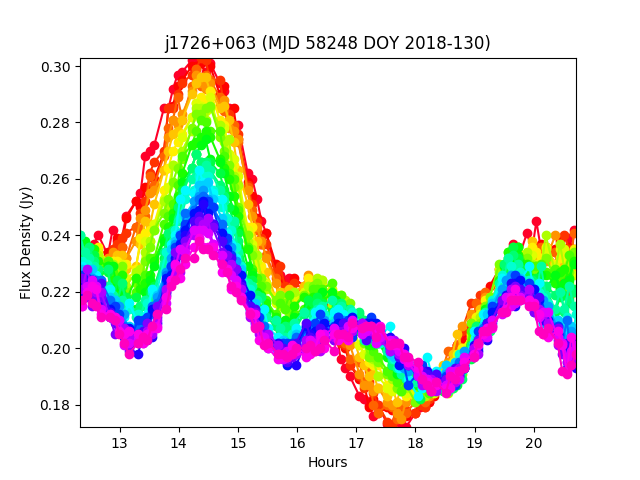}
	\includegraphics[width=\columnwidth]{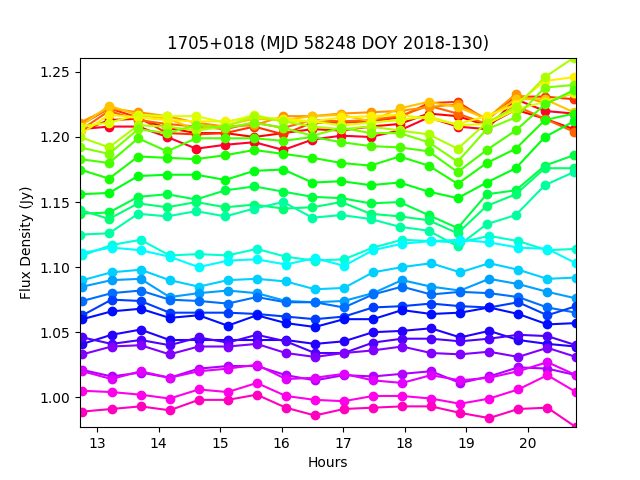}
    \caption{The top image shows light curves for PMN~J1726+0639 during an epoch of rapid scintillation, at frequencies between 4500 MHz and 10600 MHz, averaged over $\sim 100$ MHz intervals and 150\,s time intervals. The lowest frequencies are in red, and the highest are shown in violet. The lower image shows light curves from the same epoch for calibration source 1705+018, highlighting the low level of fractional systematic variations compared with the large amplitude changes observed for PMN~J1726+0639.}
    \label{fig:lightcurves_singleepoch}
\end{figure}

IHV of PMN~J1726+0639 was discovered from several short snapshots over a few hours, observed with the ATCA on two separate days in August 2017, and confirmed in November 2017. 
In order to measure an annual cycle in the rate of scintillation, and to test whether it repeated the following year, we subsequently observed PMN~J1726+0639 with the Australia Telescope Compact Array (ATCA) at 26 epochs. The epochs are not evenly spaced, as more data during the low-rate (long timescale) periods are required to best constrain the annual cycle, as discussed in \citet{Bignall19}. The observations were performed by switching between two frequency tunings; the ATCA back-end enables $2 \times 2$\,GHz-wide bands to be observed simultaneously. The resultant quasi-simultaneous spectra extend from approximately $4.3\,\mathrm{GHz}$ to $11\,\mathrm{GHz}$.
The initial flagging of radio frequency interference (RFI) and calibration of the data was done in a standard way using the Miriad software package\footnote{The calibration method is described at \url{ https://www.atnf.csiro.au/computing/software/miriad/userguide/node87.html}}, with bandpass, polarization leakage and initial gain corrections derived from the ATCA primary calibrator PKS~1934$-$638. 
The data on PMN~J1726+0639 were then split into 150\,s time slices, and each slice independently self-calibrated in Miriad. The nearby ATCA calibrator 1705+018 was observed every half hour during the monitoring runs, and each scan self-calibrated independently just as for PMN~J1726+0639, for comparison in order to check that the variations observed for PMN~J1726+0639 are not an instrumental effect. 
As PMN~J1726+0639 is isolated and unresolved with the ATCA at the observed frequencies, flux densities were estimated by averaging visibilities over all baselines. An example of the light curves obtained is shown in Figure~\ref{fig:lightcurves_singleepoch}, and for comparison the light curves obtained for the calibration source 1705+018 at the same epoch are also shown. 
For the present analysis, the data products for PMN~J1726+0639 from each epoch are between 22 and 90 spectra, of $150\,\mathrm{s}$ integration time each. The summary of these data is given in Table~\ref{table:epochs}.

\begin{table}
	\centering
	\caption{Parameters of the 26 observational epochs on which long light curves were obtained. The right column shows the number of data points remaining in the $(5.5\pm0.25)/(10\pm0.25)\,\mathrm{GHz}$ sub-bands, as used in Section~\ref{section:analysis}, after RFI excision.}
	\label{table:epochs}
\begin{tabular}{lcrrr}
	\hline
	Epoch & Date & D.o.Y. & MJD (mean) & \#points\\
	\hline
	1 & 2018/02/10 & 41 & 58159.04 & 39\\
	2 & 2018/02/11 & 42 & 58159.99 & 66/65 \\
	3 & 2018/02/24 & 54 & 58172.93 & 54/57 \\
	4 & 2018/04/05 & 94 & 58212.74 & 84/83 \\
	5 & 2018/05/11 & 130 & 58248.69 & 90 \\
    6 & 2018/07/23 & 203 & 58321.53 & 71\\
    7 & 2018/08/12 & 224 & 58342.36 & 60/59\\
    8 & 2018/08/18 & 230 & 58348.29 & 45/43 \\
    9 & 2018/08/26 & 238 & 58356.38 & 54/55\\
    10& 2018/08/27 & 239 & 58357.45 & 66/65\\
    11& 2018/10/07 & 280 & 58398.25 & 56 \\
    12& 2018/12/09 & 343 & 58461.09 & 86/85 \\
    13& 2019/03/31 & 89(+365) &58572.8 & 88\\
    14& 2019/04/08 & 97(+365) &58580.74 & 80\\
    15& 2019/05/30 & 149(+365) &58632.62 & 85\\
    16& 2019/08/19 & 231(+365) &58714.4 & 79/80\\
    17& 2019/08/27 & 239(+365) &58722.36 & 81/82\\
    18& 2019/09/01 & 244(+365) &58727.39 & 75/74\\
    19& 2019/09/06 & 249(+365) &58732.48 & 22\\
    20& 2019/09/13 & 256(+365) &58739.35 & 80/79\\
    21& 2019/09/22 & 265(+365) &58748.28 & 84\\
    22& 2019/10/07 & 280(+365) &58763.28 & 73/70\\
    23& 2019/10/26 & 299(+365) &58782.27 & 53\\
    24& 2019/12/08 & 342(+365) &58825.11 & 82/81\\
    25& 2020/03/10 & 69(+731) &58917.83 & 82/81\\
    26& 2020/03/29 & 88(+731) &58936.79 & 83/82\\
	\hline
\end{tabular}
\end{table}

\begin{figure*}
	\includegraphics[width=0.9\linewidth]{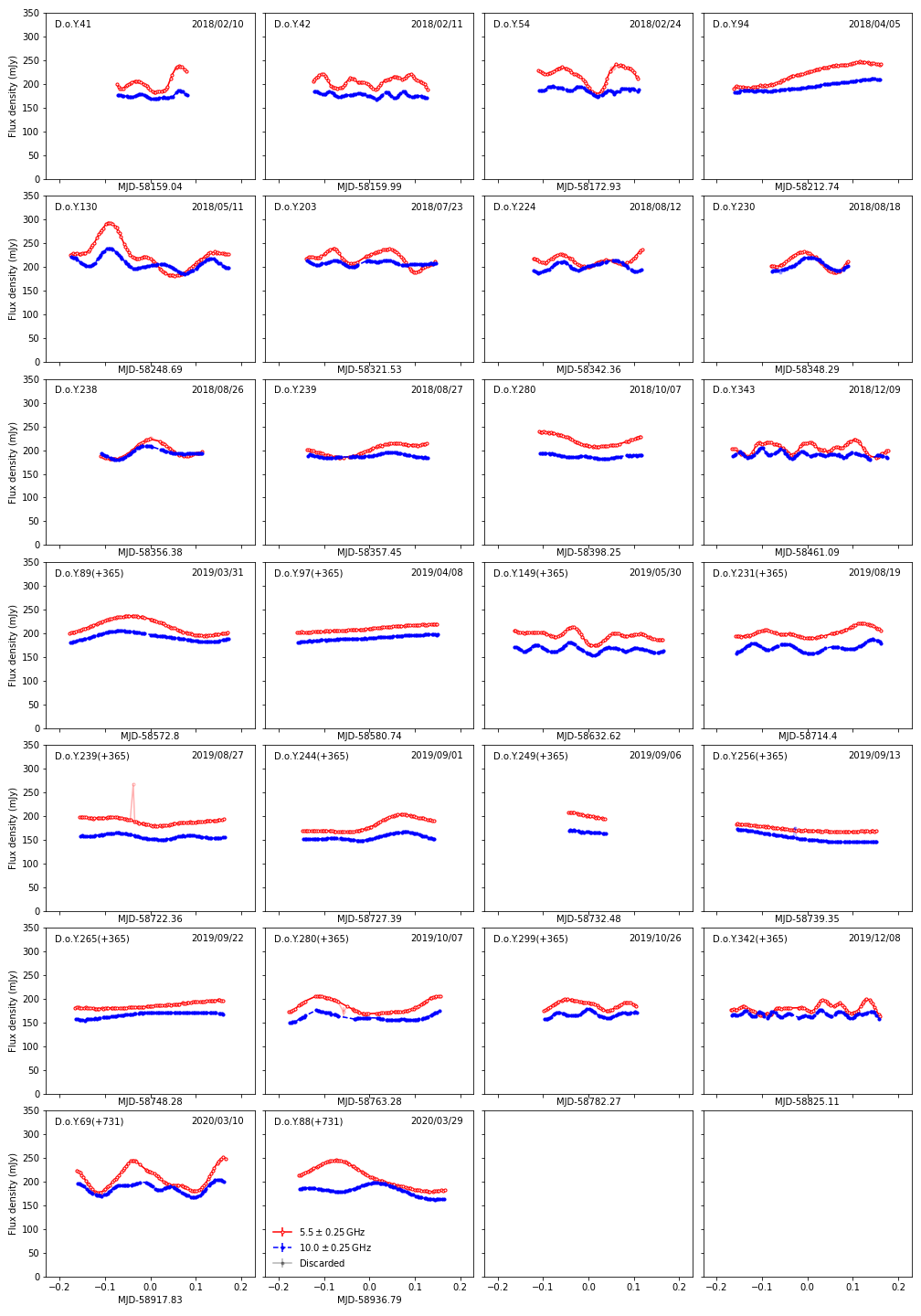}
    \caption{Observed light curves of PMN~J1726+0639, averaged over $0.5\,\mathrm{GHz}$-wide bands near the bottom and top of our bandwidth. The error bars, mostly too small to be seen, estimate the uncertainty of the mean. The points shown in semi-transparent colour were discarded from the analysis.}
    \label{figure:lightcurves}
\end{figure*}

To form the light curves used in the variability rate analysis below, we first filtered outliers (mostly due to residual unflagged radio-frequency interference, RFI) from each recorded spectrum in the sub-bands of interest, by discarding data points that deviated from the mean of the group of their 10 closest neighbours by more than 3 times the r.m.s.\ values of the group, repeating this procedure twice on the updated spectra. We then visually inspected the full dynamic spectra and dropped those remaining data points that were clearly affected by RFI or other instrumental issues. 
Figure~\ref{figure:lightcurves} presents the light curves of PMN~J1726+0639 observed at all 26 epochs, averaged over two $0.5\,\mathrm{GHz}$-wide bands centred at $5.5\,\mathrm{GHz}$ and $10\,\mathrm{GHz}$.

\section{Analysis of variability rates and the annual cycle}
\label{section:analysis}

\subsection{Scintillation rate measurement}

The novel method of Bayesian analysis used for inferring the rate of scintillation as a function of time from the light curves is described in detail in \citet{Bignall19}, and not reproduced here; the method was validated using an independent implementation in~\cite{Oosterloo20}. Briefly, using the {\tt celerite\/} package \citep{celerite} we evaluated the scintillation rate for each epoch, by determining a scaling factor for a temporal autocorrelation function whose shape is assumed constant over the year. As discussed in \citet{Bignall19}, a major advantage of this method over traditional methods to estimate characteristic variability timescale is its ability to constrain the rate near "standstill" periods, where the lightcurves show essentially no variability due to the relative velocity component of the scintillation pattern with respect to the observer being zero in the direction perpendicular to the elongation of the pattern, for a highly anisotropic scintillation pattern. Accurate determination of the zero-crossing times, or standstills, if present, in the annual cycle is crucial to determine the kinematic parameters ({\it cf.} Figure~\ref{figure:hodograph}). 

Although we have detailed spectral information as shown in Figure~\ref{fig:lightcurves_singleepoch}, the variations are partially correlated across the entire band. Based on the large amplitude of the variations, up to $\sim 50$\%, we are likely observing at frequencies close to the transition between weak and strong scattering, rather than in asymptotically weak scattering. Furthermore, the smoothness of the variations suggests the finite source size is smoothing over the scintillation pattern, thus the underlying amplitude for scintillation of a point source would be higher still. The source size, its flux density and core dominance, as well as the scintillation itself are all frequency dependent. In order to avoid taking a computationally expensive account of inter-channel correlations, while still using the information at different frequencies, as was done for \citet{Bignall19} we selected only two, widely separated $0.5\,\mathrm{GHz}$-wide bands, centred at $5.5\,\mathrm{GHz}$ and $10\,\mathrm{GHz}$ for the present analysis. These are assumed independent, and $0.5\,\mathrm{GHz}$ is chosen as the width where the empirical uncertainty of the mean over the interval (which includes both noise and real variations with frequency) approaches the expected thermal noise in the interval. This value is $\sim0.3\,\mathrm{mJy}$ for both sub-bands.

\subsection{Modelling the annual rate curve}

As described in \citet{Bignall19}, for a highly anisotropic scintillation pattern, the absolute rate of the variations,  \bea\label{absoluterate}
R=\frac{v^\perp_\mathrm{eff}}{a_\perp},
\eea
is determined by the effective transverse velocity \citep{effectivevelocity},
\bea\label{veff}
{\bf v}_\mathrm{eff}={\bf v}_\mathrm{screen}-{\bf v}_\oplus,
\eea
and $a_\perp$, defined as the half-width at half-maximum (HWHM) of the auto-correlation function (ACF) of the spatial structure of the scintillation pattern. The infinitely anisotropic (1D) model has three parameters: the position angle of anisotropy in the scintillation pattern, $\mathrm{PA}$; the transverse scattering screen velocity component perpendicular to that elongation, $v^{\perp}$; and, the characteristic size of the pattern $a_\perp$.

For the present analysis, we also extend our fitting to a more general, finitely anisotropic (2D) model of the scintillation rate, as described in \citet{Wang21}, by performing a grid search over the parameter space to minimise $\chi^2$. While the data, with clear standstill points in the annual cycle in variation rate, suggest a highly anisotropic scintillation pattern, the 2D model allows us to constrain the degree of anisotropy.   
In this model we assume that the statistics of the light curves can be described as a Gaussian process, and we have have five free parameters in the fit: 
two measurements of the spatial auto-covariance ellipse of the projected flux pattern along its principal axes $a_{\parallel,\perp}$; the orientation of the major axis of this pattern, $\mathrm{PA}$; and, two components $v^\parallel_\mathrm{screen},v^\perp_\mathrm{screen}$ of the projected screen velocity, as per
\begin{eqnarray}\label{ratesquared}
R_i^2=\frac{\left(v^\parallel_{\oplus, i}-v^\parallel_\mathrm{screen}\right)^2}{a_\parallel^2}+\frac{\left(v^\perp_{\oplus,i}-v^\perp_\mathrm{screen}\right)^2}{a_\perp^2},
\end{eqnarray}
where ${\bf v}_{\oplus,i}$ is the Earth velocity on $i$-th epoch. We minimise the $\chi^2$ formed with $R_i^2$ rather than $R_i$ because this yields an (almost, subject to the positivity constraint) linear optimisation problem in $a_{\perp, \parallel}^{-2}$, which we solve analytically in every cell of the grid; the remaining three dimensions ($\mathrm{PA}$ and $v_\mathrm{screen}^\perp, v_\mathrm{screen}^\parallel$) are feasible to explore numerically.

\subsection{Results}

\begin{figure}
	\includegraphics[width=\columnwidth]{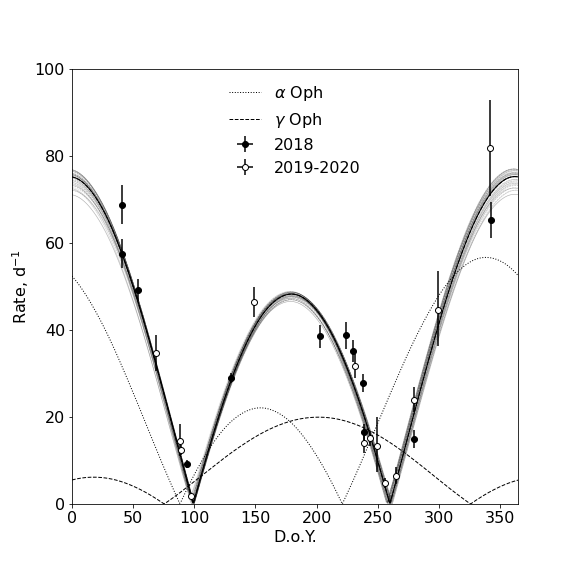}
    \caption{Absolute variation rates versus day of the year, at $5.5\,\mathrm{GHz}$.
    Models using the predicted kinematic parameters from the model of \citet{walkeretal2017} for the two nearby hot stars are shown as dotted and dashed lines; both are excluded by the data. Also shown is a selection of viable (having a $\chi^2$ within its expected 68 per cent deviation from the expected minimum) models from a grid search including finitely anisotropic scintillation patterns. The best fit model (which is totally anisotropic, 1D) is shown with a thick line.}
    \label{figure:rates-predictions}
\end{figure}

Figure~\ref{figure:rates-predictions} presents the inferred absolute rates of scintillation as a function of day of year compared to model rate curves expected for radially elongated plasma associated with hot stars along the line of sight to PMN~J$1726+0639$, as well as results of unrestricted modelling. Rates estimated at $5.5\,\mathrm{GHz}$ are shown, absolute rates at $10\,\mathrm{GHz}$ are higher by a factor of $1.49\,\pm0.01$ (the relative rates at the two frequencies are the same by construction).
We choose to plot the scintillation rates rather than timescales because, for a one-dimensional model, the former are proportional to a component of the effective velocity, $v^\perp_\mathrm{eff}$ and are thus expected to be a sinusoidal function of time.

\begin{table}
	\centering
	\caption{Parameters of acceptable fits to the observed annual cycle in the scintillation rate for PMN~J1726+0639.}
	\label{table:modellingresults}
\begin{tabular}{ll}
    \hline
	$\mathrm{PA}$ & $333^\circ\pm3^\circ$\\
	$v_\perp$ & $(5.3\pm0.8)\,\mathrm{km}\,\mathrm{s}^{-1}$\\
	$a_\perp$ & $(3.73\pm0.03)\times10^4\,\mathrm{km}$\\
	$a_\parallel: a_\perp$ & $>13:1$ (best fit is 1D)\\
	$v_\parallel$ & unconstrained (within $\pm50\,\mathrm{km}\,\mathrm{s}^{-1}$)\\
	\hline
\end{tabular}
\end{table}

\begin{figure}
	\includegraphics[width=\columnwidth]{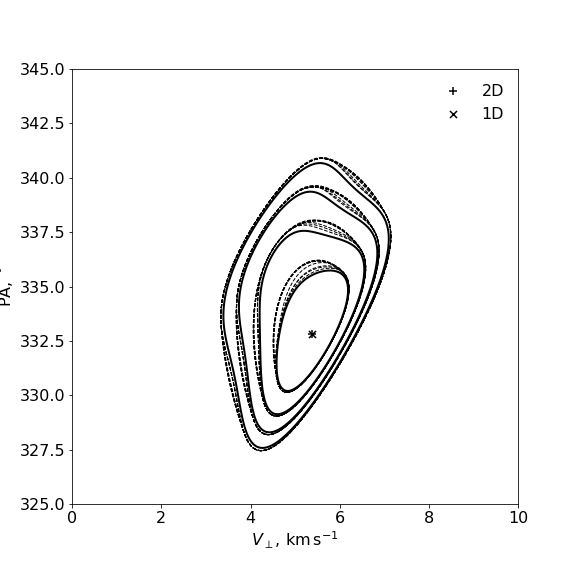}
    \caption{Posterior distribution of the parameters of the anisotropic scattering screen, contours drawn at the levels of one, two, three and four expected deviations of $\chi^2$ from its minima. $\mathrm{PA}$ is the orientation of the major axis of the scintillation pattern ($\mathrm{N}\to\mathrm{E})$ and $v_\perp$ is the velocity along the minor axis (in the direction of $\mathrm{PA}+90^\circ$). Purely 1D model contours are shown in solid lines whereas finitely anisotropic models are represented by a bunch of dashed lines (very weakly) dependent on the assumed velocity along the major axis spanning a $\pm50\,\mathrm{km}\,\mathrm{s}^{-1}$ range.}
    \label{figure:kinematics}
\end{figure}

Table~\ref{table:modellingresults} summarises the results of our modelling. The data are consistent with only highly anisotropic scintillation, with the ratio $a_\parallel:a_\perp$ constrained to be above $13:1$ for a good fit to the data. Formally the best fit model is infinitely anisotropic\footnote{We form $\chi^2$ to be minimised using $R_i^2$, rather than $R_i$, values (equation~\ref{ratesquared}), solving a linear minimisation problem for $a_{\perp, \parallel}^{-2}$ at each cell in the grid search; since $a^{-2}$ are restricted to be positive, the optimum can be reached at the boundary, $a^{-2} \to 0$, which corresponds to infinitely anisotropic models.}. 
Figure~\ref{figure:kinematics} presents the results of the modelling, reparameterised back to the $(v_\mathrm{screen}^\perp, \mathrm{PA})$ space. The parameters are well constrained with $\mathrm{PA} = 333^\circ \pm 3^{\circ}$ (north through east) and $v_\mathrm{screen}^\perp = 5.3 \pm 0.8\,\mathrm{km\,s}^{-1}$. The fitted value for the linear scale of the scintillation pattern, $a_\perp = \left(3.73\pm0.03\right) \times 10^4\,\mathrm{km}$ at $5.5\,\mathrm{GHz}$.

As discussed in \citet{Bignall19} for the case of variations observed for PKS~B1322$-$110, in \citet{Wang21} for ASKAP fast scintillators,  and also found here for the analysis of PMN~J1726+0639, errors in the fitted rates formally estimated from the Markov chain Monte Carlo (MCMC) sampling result in best fit $\chi^2$ values being much larger than expected. The errors are most likely being underestimated due to our statistical model not being entirely adequate. The one-sigma confidence region shown in Figure~\ref{figure:kinematics} corresponds to error bars inflated by a factor of 5.4 in order to bring the resultant optimal $\chi^2$ to its expected value, i.e.\ a reduced $\chi^2$ of 1, for the best fit finitely anisotropic model. Without this inflation the uncertainties are tiny. We note that our algorithm treats the two light curves as independent, but since the variations in the two bands are partially correlated, the algorithm assigns more confidence to the results than it should. We repeated the analysis using single-band light curves, but the error bars still had to be inflated by a (smaller) factor of 4 to bring the $\chi^2$ to its expected value, so the correlated light curves are not the major factor responsible for the underestimated errors. As discussed in \citet{Bignall19} the assumption of a Gaussian process and the particular parameterisation used for the auto-correlation function may not be an entirely adequate statistical description of the process, but at present we do not have a better choice.

\section{Discussion}
\label{section:discussion}

The fitted value of $a_\perp\approx 3.7 \times 10^4\,\mathrm{km}$ corresponds to a distance $D$ of the scattering screen of $\sim 5\,\mathrm{pc}$ if $a_\perp$ is interpreted as the size of the first Fresnel zone $r_{\mathrm{F}} = \sqrt{cD/(2\pi \nu)}$, as would be appropriate for weak scattering \citep[e.g.][]{Walker98}. Alternatively, following \cite{walkeretal2017}, we might interpret the observed scintillation pattern size as the projection of a $\theta_\perp=c/\nu\sqrt{F_\nu/(2\pi k_B T_B)}\approx 17\,\mathrm{\mu as}$-wide Gaussian source (times $2\sqrt{\log2}$ as appropriate for the ACF HWHM measure) implied by the observed flux density $F_\nu\approx200\,\mathrm{mJy}$ and an assumed brightness temperature of $T_B=10^{13}\,\mathrm{K}$. In that case the distance is estimated as $\sim 9\,\mathrm{pc}$. For PMN~J1726+0639\ the measured ratio of the time scales at the two frequencies is $1.49\pm0.01$. That is closer to the expected ratio of Fresnel scales ($1.35$) than the ratio of source sizes ($1.92$, assuming a constant brightness temperature), so we prefer to interpret $a_\perp$ as the size of the Fresnel zone. We note, however, that PMN~J1726+0639\ shows rms variations of $\sim 10$\%, indicating that the scintillation at 5~GHz on this line of sight is not very far from strong scattering.

As shown in Figure~\ref{figure:rates-predictions}, the annual cycle observed for J1763+063 is not consistent with expectations from the model of \citet{walkeretal2017}, assuming scattering associated with radially oriented plasma filaments around either of the nearby hot stars $\alpha\,\mathrm{Oph}$, at a distance of $14.9 \pm 0.2\,\mathrm{pc}$, or $\gamma\,\mathrm{Oph}$, at $31.5 \pm 0.2\,\mathrm{pc}$. There are no other candidate hot stars within $100\,\mathrm{pc}$ having an impact parameter of less than $3\,\mathrm{pc}$.\footnote{Just a little further out, at a distance $\simeq 110\,\mathrm{pc}$, there is the triple star system $\alpha\,\mathrm{Her}$ which incorporates one hot star. Our derived kinematic parameters $(v_\mathrm{screen}^\perp, \mathrm{PA})$ are approximately as expected for $\alpha\,\mathrm{Her}$ in the \citet{walkeretal2017} picture. However, the large distance of the stellar system makes for an implausible association in this case.}

\begin{figure}
	\includegraphics[width=\columnwidth]{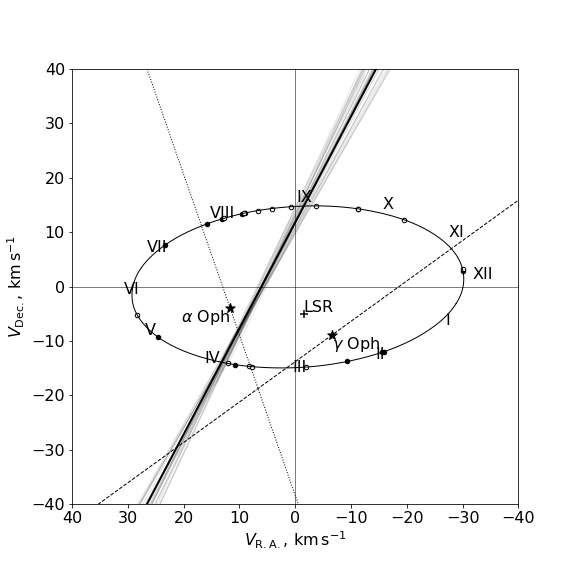}
    \caption{The hodograph for PMN~J1726+0639. The ellipse is the projection of the Earth's orbital velocity onto the plane orthogonal to the line of sight to the source. Dots along the ellipse mark the observing epochs; Roman numerals mark the mid-points of the respective months. In the hodograph, highly anisotropic scintillation models can be characterised by straight lines. We show the \citet{walkeretal2017} predictions for plasma associated with $\alpha$ and $\gamma$~Oph as dotted and dashed lines, respectively. Those predictions clearly do not match our data: we show the best-fit 1D model as a solid black line, surrounded by solid grey lines that illustrate a selection of models lying within the inner contour of figure~\ref{figure:kinematics}. The projection of the local standard of rest (LSR) is marked with a cross. }
    \label{figure:hodograph}
\end{figure}

It is instructive to represent this modelling in a hodograph plot -- a representation that was introduced in \cite{Bignall19} -- as shown in Figure~\ref{figure:hodograph}. On this diagram 1D models can be represented by straight lines with a position angle equal to that of the major axis of the flux pattern, and displaced from the origin by $v_\perp$ (in the direction $\mathrm{PA}+90^\circ$). The rate predicted for a given date is given by the perpendicular distance to this line from the corresponding date on the hodograph ellipse of the Earth's orbital motion. In particular, the rate is zero where the model crosses the orbit and, conversely, the positions of the two standstills determine the orientation and origin of the corresponding model line. The \citet{walkeretal2017} models corresponding to $\alpha$ and $\gamma$ Oph cross the orbit at positions far from the observed standstills and are thus ruled out. Comparing our figure~\ref{figure:hodograph} with figure 5 of \cite{Bignall19} makes it clear why the low ecliptic latitude of PKS~B1322-110 causes problems: when the orbit projects to a very thin ellipse, even a small uncertainty in measuring the time of the standstills corresponds to a large uncertainty in the derived position angle.

Our results for the annual cycle of PMN~J1726+0639\ do not support the model suggested by \citet{walkeretal2017}, but nor do they rule it out: it remains true that there is a low probability of a chance coincidence between hot, nearby stars and the kinematic characteristics displayed by the plasma responsible for the scintillations of J1819+3845, PKS~1257$-$326 and PKS~B1322$-$110. Similarly, the B3 star Alkaid is surprisingly close to the recently discovered IHV source J1402+5347, with a chance coincidence probability of only $\sim 2\times 10^{-3}$ for this pairing taken in isolation \citep[][]{Oosterloo20}. On the other hand there is no hot star that can plausibly explain the group of IHV sources found by \citet{Wang21}. We conclude, therefore, that although some of the plasma responsible for IHV is likely to be associated with hot stars, the overall picture of the IHV phenomenon must have more to it than just that.

The observations of \citet{Wang21} revealed 5 IHV sources in a 2-degree long linear arrangement on the sky, and those authors noted that similar structures could be abundant in the solar neighbourhood. With that result in mind it is interesting to note that there is spectroscopic evidence for a local plasma cloud in the vicinity of the PMN~J1726+0639\ line-of-sight. Specifically: there is a strong Ca~{\sc ii} (and weak Na~{\sc i}) absorption seen at an unusually high radial velocity ($-26\;{\rm km\,s^{-1}}$) towards $\alpha$~Oph (Rasalhague) \citep[][and references therein]{Redfield07}. High ratios of Ca~{\sc ii}~:~Na~{\sc i} equivalent widths are characteristic of the ``fast'' interstellar cloud population \citep[e.g.][]{SilukSilk1974}. In order to be foreground to $\alpha$~Oph this particular plasma cloud must be closer than $14.9\,\mathrm{pc}$. The same cloud was also detected by \citet{Redfield07} in the more distant star HR~6594, which lies at an angular separation of about 4 degrees from Rasalhague, demonstrating that the absorber is interstellar and not circumstellar. Based on the lack of absorption features observed towards several other (more distant) stars within a few degrees of $\alpha$~Oph, \citet{Redfield07} speculated that this intervening plasma cloud might have a filamentary morphology. Interestingly, PMN~J1726+0639\ lies only 6 degrees from $\alpha$~Oph, in a direction roughly opposite to that of HR~6594, so it is conceivable that a single, elongated plasma cloud could be responsible for both the scintillations of PMN~J1726+0639\ and the absorption lines detected towards the two stars. Motivated by the discovery of a population of H~{\sc i} absorbers that are kinematically similar to the fast plasma clouds seen in Ca~{\sc ii} \citep{Dwarakanath04,Mohan2004}, we are conducting further observations to search for an H~{\sc i} absorption counterpart towards PMN~J1726+0639.

\section{Conclusions}
\label{section:conclusions}
Monitoring of PMN~J1726+0639 has revealed a strong and repeated annual cycle in the rate of its scintillations. The cycle, which includes two standstills, is consistent with a highly anisotropic model of the scintillation. The position angle of the anisotropy, and the component of transverse scattering screen velocity perpendicular to that position angle, are well determined from the annual cycle. Those determinations show that the scintillation of PMN~J1726+0639 is not caused by radially oriented plasma filaments around a nearby hot star, as has been previously suggested for three other IHV sources. We speculate that it may instead be associated with a known, local plasma cloud (within $\sim 15\,\mathrm{pc}$) that has been previously discovered spectroscopically.  Observations with ASKAP over the next few years are expected to reveal many more, and fainter rapidly scintillating sources, offering the possibility to map out local interstellar plasma filaments. Measuring annual cycles allows the kinematics of the scattering plasma to be determined, so that coherent structures responsible for scattering in the solar neighbourhood could be identified and potentially associated with structures seen at other wavelengths. 

\section*{Acknowledgements}
 The Australia Telescope Compact Array is part of the Australia Telescope National Facility (grid.421683.a) which is funded by the Australian Government for operation as a National Facility managed by CSIRO. We acknowledge the Gomeroi people as the traditional owners of the Observatory site.
 Observations reported in this paper were made under ATCA project code C3214.


\section*{Data availability}
The data underlying this article are available in the Australia Telescope Online Archive at  https://atoa.atnf.csiro.au/ and can be accessed under project code C3214. The derived data generated in this research will be shared on reasonable request to the corresponding author.

\bibliographystyle{mnras}
\bibliography{J1726cycle} 

\begin{thebibliography}{}
\makeatletter
\relax
\def\mn@urlcharsother{\let\do\@makeother \do\$\do\&\do\#\do\^\do\_\do\%\do\~}
\def\mn@doi{\begingroup\mn@urlcharsother \@ifnextchar [ {\mn@doi@}
  {\mn@doi@[]}}
\def\mn@doi@[#1]#2{\def\@tempa{#1}\ifx\@tempa\@empty \href
  {http://dx.doi.org/#2} {doi:#2}\else \href {http://dx.doi.org/#2} {#1}\fi
  \endgroup}
\def\mn@eprint#1#2{\mn@eprint@#1:#2::\@nil}
\def\mn@eprint@arXiv#1{\href {http://arxiv.org/abs/#1} {{\tt arXiv:#1}}}
\def\mn@eprint@dblp#1{\href {http://dblp.uni-trier.de/rec/bibtex/#1.xml}
  {dblp:#1}}
\def\mn@eprint@#1:#2:#3:#4\@nil{\def\@tempa {#1}\def\@tempb {#2}\def\@tempc
  {#3}\ifx \@tempc \@empty \let \@tempc \@tempb \let \@tempb \@tempa \fi \ifx
  \@tempb \@empty \def\@tempb {arXiv}\fi \@ifundefined
  {mn@eprint@\@tempb}{\@tempb:\@tempc}{\expandafter \expandafter \csname
  mn@eprint@\@tempb\endcsname \expandafter{\@tempc}}}

\bibitem[\protect\citeauthoryear{{Bignall} et~al.,}{{Bignall}
  et~al.}{2019}]{Bignall19}
{Bignall} H.,  et~al., 2019, \mn@doi [\mnras] {10.1093/mnras/stz1559}, \href
  {https://ui.adsabs.harvard.edu/abs/2019MNRAS.487.4372B} {487, 4372}

\bibitem[\protect\citeauthoryear{{Cordes} \& {Rickett}}{{Cordes} \&
  {Rickett}}{1998}]{effectivevelocity}
{Cordes} J.~M.,  {Rickett} B.~J.,  1998, \mn@doi [\apj] {10.1086/306358}, \href
  {http://adsabs.harvard.edu/abs/1998ApJ...507..846C} {507, 846}

\bibitem[\protect\citeauthoryear{{Dennett-Thorpe} \& {de
  Bruyn}}{{Dennett-Thorpe} \& {de Bruyn}}{2001}]{cycle1819}
{Dennett-Thorpe} J.,  {de Bruyn} A.~G.,  2001, in {Laing} R.~A.,  {Blundell}
  K.~M.,  eds,  Astronomical Society of the Pacific Conference Series Vol. 250,
  Particles and Fields in Radio Galaxies Conference. p.~133

\bibitem[\protect\citeauthoryear{{Dennett-Thorpe} \& {de
  Bruyn}}{{Dennett-Thorpe} \& {de Bruyn}}{2002}]{twostation1819}
{Dennett-Thorpe} J.,  {de Bruyn} A.~G.,  2002, \mn@doi [\nat]
  {10.1038/415057a}, \href {http://adsabs.harvard.edu/abs/2002Natur.415...57D}
  {415, 57}

\bibitem[\protect\citeauthoryear{{Dwarakanath}}{{Dwarakanath}}{2004}]{Dwarakanath04}
{Dwarakanath} K.~S.,  2004, Bulletin of the Astronomical Society of India,
  \href {https://ui.adsabs.harvard.edu/abs/2004BASI...32..215D} {32, 215}

\bibitem[\protect\citeauthoryear{Foreman-Mackey, Agol, Ambikasaran  \&
  Angus}{Foreman-Mackey et~al.}{2017}]{celerite}
Foreman-Mackey D.,  Agol E.,  Ambikasaran S.,   Angus R.,  2017, The
  Astronomical Journal, 154, 220

\bibitem[\protect\citeauthoryear{{Griffith}, {Wright}, {Burke}  \&
  {Ekers}}{{Griffith} et~al.}{1995}]{Griffith95}
{Griffith} M.~R.,  {Wright} A.~E.,  {Burke} B.~F.,   {Ekers} R.~D.,  1995,
  \mn@doi [\apjs] {10.1086/192146}, \href
  {https://ui.adsabs.harvard.edu/abs/1995ApJS...97..347G} {97, 347}

\bibitem[\protect\citeauthoryear{{Jauncey} \& {Macquart}}{{Jauncey} \&
  {Macquart}}{2001}]{jaunceymacquart}
{Jauncey} D.~L.,  {Macquart} J.-P.,  2001, \mn@doi [\aap]
  {10.1051/0004-6361:20010299}, \href
  {http://adsabs.harvard.edu/abs/2001A%26A...370L...9J} {370, L9}

\bibitem[\protect\citeauthoryear{{Lovell} et~al.,}{{Lovell}
  et~al.}{2008}]{Lovell08}
{Lovell} J.~E.~J.,  et~al., 2008, \mn@doi [\apj] {10.1086/592485}, \href
  {https://ui.adsabs.harvard.edu/abs/2008ApJ...689..108L} {689, 108}

\bibitem[\protect\citeauthoryear{{Mohan}, {Dwarakanath}  \&
  {Srinivasan}}{{Mohan} et~al.}{2004}]{Mohan2004}
{Mohan} R.,  {Dwarakanath} K.~S.,   {Srinivasan} G.,  2004, \mn@doi [Journal of
  Astrophysics and Astronomy] {10.1007/BF02702371}, \href
  {https://ui.adsabs.harvard.edu/abs/2004JApA...25..185M} {25, 185}

\bibitem[\protect\citeauthoryear{{Oosterloo} et~al.,}{{Oosterloo}
  et~al.}{2020}]{Oosterloo20}
{Oosterloo} T.~A.,  et~al., 2020, \mn@doi [\aap] {10.1051/0004-6361/202038378},
  \href {https://ui.adsabs.harvard.edu/abs/2020A&A...641L...4O} {641, L4}

\bibitem[\protect\citeauthoryear{{Redfield}, {Kessler-Silacci}  \&
  {Cieza}}{{Redfield} et~al.}{2007}]{Redfield07}
{Redfield} S.,  {Kessler-Silacci} J.~E.,   {Cieza} L.~A.,  2007, \mn@doi [\apj]
  {10.1086/517516}, \href
  {https://ui.adsabs.harvard.edu/abs/2007ApJ...661..944R} {661, 944}

\bibitem[\protect\citeauthoryear{{Rickett}, {Witzel}, {Kraus}, {Krichbaum}  \&
  {Qian}}{{Rickett} et~al.}{2001}]{rickettetal}
{Rickett} B.~J.,  {Witzel} A.,  {Kraus} A.,  {Krichbaum} T.~P.,   {Qian} S.~J.,
   2001, \mn@doi [\apjl] {10.1086/319493}, \href
  {http://adsabs.harvard.edu/abs/2001ApJ...550L..11R} {550, L11}

\bibitem[\protect\citeauthoryear{{Said}, {Ellingsen}, {Bignall}, {Shabala},
  {McCallum}  \& {Reynolds}}{{Said} et~al.}{2020}]{Said20}
{Said} N.~M.~M.,  {Ellingsen} S.~P.,  {Bignall} H.~E.,  {Shabala} S.,
  {McCallum} J.~N.,   {Reynolds} C.,  2020, \mn@doi [\mnras]
  {10.1093/mnras/staa2642}, \href
  {https://ui.adsabs.harvard.edu/abs/2020MNRAS.498.4615S} {498, 4615}

\bibitem[\protect\citeauthoryear{{Siluk} \& {Silk}}{{Siluk} \&
  {Silk}}{1974}]{SilukSilk1974}
{Siluk} R.~S.,  {Silk} J.,  1974, \mn@doi [\apj] {10.1086/153033}, \href
  {https://ui.adsabs.harvard.edu/abs/1974ApJ...192...51S} {192, 51}

\bibitem[\protect\citeauthoryear{{Walker}}{{Walker}}{1998}]{Walker98}
{Walker} M.~A.,  1998, \mn@doi [\mnras] {10.1046/j.1365-8711.1998.01238.x},
  \href {https://ui.adsabs.harvard.edu/abs/1998MNRAS.294..307W} {294, 307}

\bibitem[\protect\citeauthoryear{{Walker}, {de Bruyn}  \& {Bignall}}{{Walker}
  et~al.}{2009}]{walkerdebruynbignall}
{Walker} M.~A.,  {de Bruyn} A.~G.,   {Bignall} H.~E.,  2009, \mn@doi [\mnras]
  {10.1111/j.1365-2966.2009.14942.x}, \href
  {http://adsabs.harvard.edu/abs/2009MNRAS.397..447W} {397, 447}

\bibitem[\protect\citeauthoryear{{Walker}, {Tuntsov}, {Bignall}, {Reynolds},
  {Bannister}, {Johnston}, {Stevens}  \& {Ravi}}{{Walker}
  et~al.}{2017}]{walkeretal2017}
{Walker} M.~A.,  {Tuntsov} A.~V.,  {Bignall} H.,  {Reynolds} C.,  {Bannister}
  K.~W.,  {Johnston} S.,  {Stevens} J.,   {Ravi} V.,  2017, \mn@doi [\apj]
  {10.3847/1538-4357/aa705c}, \href
  {http://adsabs.harvard.edu/abs/2017ApJ...843...15W} {843, 15}

\bibitem[\protect\citeauthoryear{{Wang}, {Tuntsov}, {Murphy}, {Lenc}, {Walker},
  {Bannister}, {Kaplan}  \& {Mahony}}{{Wang} et~al.}{2021}]{Wang21}
{Wang} Y.,  {Tuntsov} A.,  {Murphy} T.,  {Lenc} E.,  {Walker} M.,  {Bannister}
  K.,  {Kaplan} D.~L.,   {Mahony} E.~K.,  2021, \mn@doi [\mnras]
  {10.1093/mnras/stab139}, \href
  {https://ui.adsabs.harvard.edu/abs/2021MNRAS.502.3294W} {502, 3294}

\makeatother
\end{thebibliography}





\bsp	
\label{lastpage}
\end{document}